\documentclass[aps,prd,
preprintnumbers,nofootinbib,superscriptaddress]{revtex4}
\usepackage[dvips]{graphicx}
\usepackage{caption}
\usepackage{subcaption}
\usepackage{comment}
\usepackage{hyperref}
\usepackage{bm,latexsym,amsmath,amssymb,amsfonts,color}

\begin{document}

\title{Self-gravitating baryonic tubes supported by $\pi$- and $\omega$-mesons \\ and its flat limit}
\author{Gonzalo Barriga}
\email{gobarriga@udec.cl}
\affiliation{Physique Théorique et Mathématique and International Solvay Institutes, Université Libre de Bruxelles (ULB), C.P. 231, 1050 Bruxelles, Belgium}
\affiliation{Departamento de F\'{\i}sica, Universidad de Concepci\'{o}n,
Casilla 160-C, Concepci\'{o}n, Chile}
\author{Carla Henr\'iquez Baez}   
\email{carla.henriquez@umayor.cl}
\affiliation{Centro Multidisciplinario de F\'isica, Vicerrector\'ia de Investigaci\'on, Universidad Mayor, Camino La Pir\'amide 5750, Santiago, Chile.}
\affiliation{Grupo de Investigación en Física Teórica, GIFT, Universidad Católica de la Santísima Concepción, Alonso de Ribera 2850, Concepción, Chile.}
\author{Leonardo Sanhueza}
\email{lsanhueza@udec.cl}
\affiliation{Departamento de F\'{\i}sica, Universidad de Concepci\'{o}n,
Casilla 160-C, Concepci\'{o}n, Chile}
\affiliation{Perimeter Institute for Theoretical Physics, 31 Caroline Street North, Waterloo, Ontario N2L 2Y5, Canada}
\author{Aldo Vera}
\email{aldo.vera@umayor.cl}
\affiliation{N\'ucleo de Matem\'atica, F\'isica y Estad\'istica, Universidad Mayor, Avenida Manuel Montt 367, Santiago, Chile}
\affiliation{Centro Multidisciplinario de F\'isica, Vicerrector\'ia de Investigaci\'on, Universidad Mayor, Camino La Pir\'amide 5750, Santiago, Chile.}

\begin{abstract}
In this paper, we construct self-gravitating topological solitons in the $SU(N)$ Einstein non-linear sigma model coupled to $\omega$-vector mesons in four space-time dimensions. These solutions represent tube-like configurations free of curvature singularities, carrying a non-vanishing topological charge that is identified as the baryon number. We show that by employing the maximal embedding Ansatz of $SU(2)$ into $SU(N)$ in the exponential representation, these tubes can be constructed for an arbitrary number of flavors, $N$, with the topological charge scaling proportionally to this number. The flat-space limit of the solutions, corresponding to an array of baryonic tubes within a finite volume, is analyzed in detail. Remarkably, while the total energy of the solitons at a finite volume is an increasing function of $N$, the binding energy decreases as the number of flavors increases. This analysis reaffirms that the inclusion of more than two flavors to the model systematically improves the physical predictions.
\end{abstract}

\maketitle

\newpage


\section{Introduction}

The non-linear sigma model (NLSM) is one of the most relevant effective field theories, originally introduced in particle physics to describe pion dynamics at low energies \cite{Nair, ChPT}. This theory has a wide range of applications that range from condensed matter physics to string theory \cite{MantonBook, BalaBook}. Despite its phenomenological success, T. Skyrme demonstrated that for the NLSM to support energetically stable topological solitons, the action must be supplemented with higher-order derivative terms to evade Derrick's scaling theorem \cite{skyr, D}. The most standard approach is the inclusion of the so-called Skyrme term. However, alternative mechanisms exist to circumvent Derrick's argument, such as coupling the NLSM to spin-$1$ vector mesons or exploring geometries beyond spherical symmetry. These mechanisms allow for the existence of configurations with a non-zero topological charge, which, within chiral effective field theories of quantum chromodynamics (QCD), is identified with the baryon number \cite{Witten1, Witten2} (see also Refs. \cite{Bala1, Bala2}).

Although the NLSM and the Skyrme model provide an accurate description of the static properties and interacting states of hadrons \cite{Hooft, Weinberg:1990rz}, experimental data reveal that these theories lack accuracy in certain relevant predictions. This discrepancy becomes particularly significant regarding the nuclear binding energy, where experimental data show that the repulsive energy between baryons is smaller than that predicted by standard chiral models. Fortunately, this issue can be resolved: hadrons can be stabilized, and the binding energy can be reduced, by introducing $\omega$-vector-mesons into the theory. This mechanism was first noticed by Adkins and Nappi in Ref. \cite{Adkins1}, where the authors reported that the presence of the $\omega$-mesons stabilizes the solitons and reduces the binding energy without the need for the Skyrme term \cite{Meissner1, Bijnens:1997ni}. Subsequently, these results were expanded to more general effective models, emphasizing the relevance of including vector mesons \cite{Park:2003sd, Leask:2024ith}.

Recently, in Ref. \cite{Barriga:2023jam}, analytical solutions of the $SU(2)$ NLSM coupled to $\omega$-mesons were constructed (see also Ref. \cite{Barriga:2022izc}). These solutions represent ordered arrays of baryonic tubes and layers with non-trivial topological charge, where the presence of $\omega$-mesons reduces the repulsive energy between baryons and leads to a kind of \textit{nuclear linguine phase} in a finite volume. These solutions were constructed using techniques previously developed in the context of the Skyrme model \cite{crystal1, Canfora:2023zmt}.

Usually, effective field theories are studied in the two-flavor case—that is, when the internal global symmetry group is $SU(2)$—in order to simplify the set of coupled non-linear differential equations. Indeed, extending the internal symmetry group to the general $SU(N)$ case yields a highly complex system of $(N^{2}-1)$ coupled partial differential equations. This complexity typically renders the extraction of analytical information for configurations with $N > 2$ almost impossible, in contrast to the $SU(2)$ case where certain exact solutions can be found. In this context, a highly effective strategy is to employ the maximal embedding Ansatz of $SU(2)$ into $SU(N)$ \cite{Bertini:2005rc, Cacciatori:2012qi}. This powerful technique drastically reduces the matter field degrees of freedom, allowing us to circumvent the non-linear intricacies and explore exactly what role the flavor number plays in the observables of the model. For instance, this formalism has recently been used to study phase transitions between nuclear pasta states in the context of the Skyrme model \cite{SU(N)1, CacciatoriScotti}.

Solutions from the effective chiral theories mentioned above can also be promoted to self-gravitating configurations by coupling the corresponding matter fields to general relativity. Indeed, relevant exact solutions in these models describing black holes, gravitating solitons, and extended objects have been successfully constructed \cite{CanforaMaeda, Canfora:2025roy}.

In this article, we construct solutions that describe self-gravitating solitons in the $SU(N)$ Einstein-NLSM coupled to $\omega$-mesons. These solutions describe regular baryonic tubes, and generalize the flat space-time solutions presented in Ref. \cite{Barriga:2023jam}. Furthermore, using the maximal embedding Ansatz in the exponential representation, we demonstrate that these solutions are robust to the inclusion of an arbitrary number of flavors. To achieve this, we will build upon the frameworks established in Refs. \cite{gravtube1, gravtube2}, where self-gravitating superconducting tubes and parallel waves in the $SU(2)$ Einstein-NLSM were constructed. Finally, we will study the flat limit of these solutions, showing that the binding energy is a monotonically decreasing function of the flavor number, underscoring that the physical predictions of the model systematically improve when considering more than two flavors.

We will demonstrate that, although the fully coupled system—particularly the $\omega$-meson sector for an arbitrary $N > 2$ in a curved background—ultimately requires the numerical integration of the governing differential equations, the highly symmetric Ansatz on the pionic sector is far from arbitrary. It is precisely this geometric simplification that allows us to extract exact analytical expressions for fundamental quantities such as the topological charge, the energy density, and the functional dependence of the binding energy on $N$. Furthermore, we will show that a completely analytical solution exists for an arbitrary number of flavors in the region where the $\omega$-mesons vanish; that is, when the tubes are supported purely by the pionic matter field. Our overarching goal is to push the analytical boundaries of hadronic tube constructions as far as possible before resorting to numerical methods.

This paper is organized as follows: In Section II, we introduce the $SU(N)$ Einstein-NLSM coupled to $\omega$-mesons. In Section III, we construct semi-analytical solutions describing self-gravitating tubes and discuss their main physical properties. In Section IV, we analyze the flat limit of our solutions and show how the inclusion of $\omega$-mesons, together with a flavor number greater than $2$, allows for a systematic reduction of the binding energy. Finally, in Section V, we present our conclusions and perspectives.

\section{General Relativity in the presence of $\pi$ and $\omega$ mesons}

\section{Field equations and the model}

As a starting point, we consider the $SU(N)$ Einstein-NLSM coupled to $\omega$-vector-mesons. The theory is described by the following action principle:
\begin{gather} \label{I}
I[g,U,\omega]=\int_{\mathcal{M}} d^{4} x \sqrt{-g}\left(\frac{\mathcal{R}}{2 \kappa}+\frac{K}{4} \operatorname{Tr}\left[L^{\mu} L_{\mu}\right]-\frac{1}{4} S_{\mu \nu} S^{\mu \nu}-\frac{1}{2} M_{\omega}^{2} \omega_{\mu} \omega^{\mu} -\gamma \rho_{\mu} \omega^{\mu}\right) , \\
U(x) \in SU(N) , \quad L_{\mu}=U^{-1} \nabla_{\mu} U , \quad S_{\mu\nu}= \partial_\mu \omega_\nu - \partial_\nu \omega_\mu , \notag
\end{gather}
where $\mathcal{R}$ is the Ricci scalar, $U(x)$ is the $SU(N)$-valued pionic scalar field, and $\omega_{\mu}$ is a four-vector representing the $\omega$-meson potential. The term $\rho_\mu$ denotes the topological current, defined in Eq.~\eqref{rhomu} below. The coupling constants are $\kappa = 8 \pi G$ (where $G$ is the gravitational constant), $M_{\omega}$ (the mass of the $\omega$-meson), and $K$, $\gamma$ which are positive constants typically fixed by experimental data. We adopt the convention $c=\hbar=1$, and Greek indices denote four-dimensional spacetime coordinates with a mostly plus signature $(-,+,+,+)$.

In this formulation, vector mesons are introduced through explicit gauge-like fields in the action, while pseudoscalar mesons emerge from the non-linear dynamics of the $SU(N)$ group. Consequently, baryons arise as topological solitons from the non-linear interactions between these mesonic sectors.

The complete set of field equations is obtained by varying the action with respect to the fields $U$, $\omega_\mu$, and the metric $g_{\mu\nu}$. These equations are given by:
\begin{gather} 
\nabla_{\mu}L^{\mu}-\frac{6 \gamma}{K}\nabla_{\nu}\left(\epsilon^{\mu\nu\lambda\rho}\omega_{\mu}L_{\lambda}L_{\rho}\right)=0 , \label{EqNLSM} \\ 
\nabla_{\mu} S^{\mu \nu}-M_{\omega}^{2} \omega^{\nu}=\gamma \rho^{\nu} , \label{Eqomega} \\ 
G_{\mu \nu}=\kappa T_{\mu \nu} , \label{EqEinstein}
\end{gather}
where $G_{\mu\nu}$ is the Einstein tensor. The $\omega$-mesons interact with the pionic sector via the topological current $\rho^{\mu}$, which is defined as:
\begin{equation} 
\rho^{\mu}=\epsilon^{\mu\nu\lambda\rho} \operatorname{Tr}\left[\left(U^{-1} \partial_{\nu} U\right)\left(U^{-1} \partial_{\lambda} U\right)\left(U^{-1} \partial_{\rho} U\right)\right] , \label{rhomu}
\end{equation}
where $\epsilon^{\mu\nu\lambda\rho}$ is the Levi-Civita tensor.\footnote{This tensor is related to the Levi-Civita symbol $\varepsilon^{\mu\nu\lambda\rho}$ and the determinant of the metric $g$ by:
\begin{equation*}
    \epsilon^{\mu\nu\lambda\rho}= - \frac{1}{\sqrt{-g}}\varepsilon^{\mu\nu\lambda\rho} , \quad \epsilon_{\mu\nu\lambda\rho}= \sqrt{-g}\varepsilon_{\mu\nu\lambda\rho} \, .
\end{equation*}}
The integral of the time-component of this current over a spacelike hypersurface $\Sigma$ yields the topological charge, interpreted as the baryon number $B$ of the system:
\begin{equation} \label{B}
    B=\frac{1}{24\pi^{2}}\int_{\Sigma}\sqrt{\sigma}n_{\mu}\rho^{\mu}d^{3}x \, . 
\end{equation}
In this expression, $n_{\mu}$ is the unit normal vector to the hypersurfaces of constant $t$, and $\sigma$ is the determinant of the induced metric on $\Sigma$. As expected for a topological quantity, the baryon number $B$ is independent of the metric, as the metric dependence in the volume element is exactly compensated by the Levi-Civita tensor (see Ref.~\cite{gravtube2} for a detailed discussion).

Finally, the energy-momentum tensor $T_{\mu\nu}$, obtained by varying the action~\eqref{I} with respect to the metric, reads:
\begin{equation}
\begin{aligned}
T_{\mu\nu} = & -\frac{K}{2} \operatorname{Tr}\left[L_\mu L_\nu-\frac{1}{2} g_{\mu \nu} L^\alpha L_\alpha\right] + S_{\mu \alpha} S_\nu{ }^\alpha - \frac{1}{4} S_{\alpha \beta} S^{\alpha \beta} g_{\mu \nu} \\
& + M_\omega^2\left(\omega_\mu \omega_\nu - \frac{1}{2} g_{\mu \nu} \omega^\alpha \omega_\alpha\right) + \gamma\left(\rho_\mu \omega_\nu + \rho_\nu \omega_\mu - g_{\mu \nu} \rho_\alpha \omega^\alpha\right) . \label{Tmunu}
\end{aligned}
\end{equation}

\section{Self-gravitating hadronic tubes}

Self-gravitating hadronic tubes can be constructed as solutions to the Einstein-NLSM by considering pionic matter field configurations with non-vanishing topological charge within a spacetime described by the Weyl-Lewis-Papapetrou (WLP) metric. This approach was first explored in Ref.~\cite{gravtube1} (see also Ref.~\cite{gravtube2}) in the context of the $SU(2)$ Einstein-NLSM coupled to Maxwell electrodynamics. In those works, the pionic matter sector leads to configurations that, in the flat-space limit, correspond to periodic arrays of baryonic tubes. It was demonstrated that such matter fields are compatible with the WLP metric, allowing the matter field equations to decouple from the Einstein equations. Moreover, by choosing an appropriate Ansatz for the gauge potential, the NLSM equations reduce to the same first-order equations found in the flat-space cases of Refs.~\cite{crystal1, crystal2}. In the present work, we adopt a similar strategy by selecting a suitable Ansatz for the $\omega$-meson potential that facilitates this decoupling.

\subsection{The Ansatz} 

In the study of the NLSM, the two-flavor case ($N=2$) is usually considered in the literature due to its relative simplicity. However, in the general $SU(N)$ case, the pionic field $U(x) \in SU(N)$ yields a system of $(N^2-1)$ coupled non-linear differential equations. This represents a formidable analytical challenge unless an appropriate Ansatz is used that allows, first, reducing the number of degrees of freedom of the system and, second, does not simply reduce to the two-flavor case.

To construct solutions where the flavor number $N$ appears explicitly in the physical observables, we employ the maximal embedding of $SU(2)$ into $SU(N)$ in the exponential representation, following the formalism developed in Refs. \cite{SU(N)1, SU(N)2}. In this representation, the pionic field is parametrized as:
\begin{gather} \label{U} 
    U = e^{\alpha(x^\mu)(\vec{n} \cdot \vec{T})} \ , \\
    \vec{n} = \left(\sin\Theta\sin\Phi, \sin\Theta \cos\Phi, \cos\Theta\right) \ , \notag    
\end{gather}
where $\alpha(x^\mu)$, $\Theta(x^\mu)$, and $\Phi(x^\mu)$ represent the three degrees of freedom, with $\alpha$ acting as the soliton profile. The matrices $T_{i}$ (with $i=1,2,3$) are the generators of a three-dimensional subalgebra of $\mathfrak{su}(N)$, explicitly given by:
\begin{align}
T_1 &= -\frac{i}{2}\sum_{j=2}^{N} \sqrt{(j-1)(N-j+1)}(E_{j-1,j}+E_{j,j-1}) \ , \label{T1} \\
T_2 &= \frac{1}{2}\sum_{j=2}^{N} \sqrt{(j-1)(N-j+1)}(E_{j-1,j}-E_{j,j-1}) \ , \label{T2} \\
T_3 &= i\sum_{j=1}^{N} \left(\frac{N+1}{2}-j\right)E_{j,j} \ , \label{T3}
\end{align}
where $(E_{i,j})_{mn}=\delta_{im}\delta_{jn}$. These generators satisfy the standard commutation relations $[T_{a}, T_{b}] = \epsilon_{abc}T_{c}$. The primary advantage of this Ansatz is that it allows the exploration of the $N$-dependence of the solutions without increasing the number of equations to be solved. 

The non-trivial nature of this embedding is manifest in the trace of the generators:
\begin{equation} \label{Tr}
\text{Tr}\left(T_{b}T_{c} \right) = - \frac{1}{2} \bar{N}\delta_{bc} \ , \qquad \bar{N} = \frac{N(N^2-1)}{6} \ . 
\end{equation}
The factor $\bar{N}$, which corresponds to the tetrahedral numbers, introduces an explicit dependence on the flavor number. As we shall see, this scaling is fundamental, as both the energy density and the topological charge are proportional to $\bar{N}$.

Following the construction of hadronic tubes in flat space-time \cite{crystal1, crystal2}, we adopt the following Ansatz for the pionic degrees of freedom:
\begin{equation}
\alpha(x^\mu) = 2\alpha(X) \ , \qquad \Theta(x^\mu) = q\theta \ , \qquad \Phi(x^\mu) = p (t-z) \ , \label{dof}
\end{equation}
where $p$ and $q$ are real parameters. We define these fields on a WLP metric:
\begin{equation} \label{metric}
ds^{2} = \left(-2+ H\right)dt^{2} +2 \left(1-H\right) dtdz + e^{-2R}(dX^{2}+d\theta^{2}) +H dz^2\ .
\end{equation}
Where $H=H(X, \theta)$ and $R=R(X)$ are metric functions with $t\in [0,\infty)$, $X \in (-\infty, \infty)$, $\theta \in [0, 2\pi)$ and $z\in (-\infty,\infty)$. Since the determinant of the $t-z$ section is everywhere negative and non-vanishing, the metric remains non-degenerate and Lorentzian, regardless of the form of the $H$ function.

For the $\omega$-meson potential, we assume:
\begin{align} \label{omega}
\omega_{\mu} = (-S(X,\theta), 0, 0, S(X,\theta)) \ ,
\end{align}
where $S$ is a scalar function. This choice of $\omega_\mu$ is specifically designed to allow the pionic equations to decouple from the vector meson equations. As shown below, this Ansatz is consistent with the stationary nature of the configuration and enables an analytical treatment of the soliton profile even in the presence of $\omega$-mesons.

\subsection{Solving the system}

Using the Ansatz presented in Eqs. \eqref{U}, \eqref{dof}, \eqref{metric}
and \eqref{omega}, the complete set of the $SU(N)$ Einstein-NLSM coupled to $\omega$-mesons is reduced in the following way: First, the NLSM equations are reduced to just one second-order ordinary differential equation for the soliton profile: 
\begin{equation} \label{Eqalpha}
    \alpha''-q^2 \cos(\alpha)\sin(\alpha)= 0 \ . 
\end{equation}
Second, the $\omega$-mesons equations are reduced to one partial differential equation for the function $S$:

\begin{equation} \label{EqS}
    \Delta S-e^{-2R}M_{\omega}^{2}S +12\bar{N}\gamma pq\sin\left(q\theta\right)\sin^{2}\left(\alpha\right) \, \alpha'=0 \ ,
\end{equation}
where $\bar{N}$ has been defined in Eq. \eqref{Tr}. On the other hand, the non-null Einstein field equations turn out to be
\begin{gather}
 (\alpha')^{2}-q^{2}\sin^{2}\left(\alpha\right)=0 \ , \label{Eqalpha0}\\
         R''-\frac{1}{2}\bar{N}K\kappa\left(q^{2}\sin^{2}\left(\alpha\right)+(\alpha')^{2}\right)=0 \ ,   \label{EqR} \\
                        \Delta H +2\kappa \left(\partial S\right)^2 -48  p q \bar{N}\gamma \kappa S \sin(q\theta)\sin^2 (\alpha)\alpha'  +2\kappa e^{-2R}\left(M_{\omega}^2 S^2 +K\bar{N}p^2 \sin^2(q \theta)\sin^2 (\alpha) \right)=0 \label{EqH} \ .
\end{gather}
Here, we are using the following notation: $\Delta \mathcal{O} = \partial^2_X \mathcal{O} + \partial^2_\theta \mathcal{O}$, and $(\partial \mathcal{O})^2 = \partial_X \mathcal{O} \partial^X \mathcal{O} + \partial_\theta \mathcal{O} \partial^\theta \mathcal{O}$.
Before starting to solve the system, it is important to note two relevant issues. First, the Einstein equation in Eq. \eqref{Eqalpha0} is the first integral of Eq. \eqref{Eqalpha} in the particular case where the integration constant that comes from Eq. \eqref{Eqalpha} vanishes. This implies that by solving Eq. \eqref{Eqalpha0}, Eq. \eqref{Eqalpha} is automatically satisfied (although the solutions of Eq. \eqref{Eqalpha} are more general than those in Eq. \eqref{Eqalpha0}). Second, also from Eq. \eqref{Eqalpha0}, we can solve $(\alpha')$ and then substitute it into the other equations in order to simplify the system.\footnote{Note that there are two possible branches for $
\alpha^\prime$ given by $\alpha^\prime=\pm q \sin(\alpha)$. Without loss of generality, we just consider the positive branch.}
  
In fact, Eqs. \eqref{EqS}, \eqref{EqR} and \eqref{EqH} become
\begin{gather}
    \Delta S-e^{-2R}M_{\omega}^{2}S+12\bar{N}\gamma pq^2\sin\left(q\theta\right)\sin^{3}\left(\alpha\right)=0 \ , \\
      R''-\bar{N}K\kappa q^{2}\sin^{2}\left(\alpha\right)=0 \ ,\\ 
      \Delta H +2\kappa \left(\partial S\right)^2 -48  p q^2 \bar{N}\gamma \kappa S \sin(q\theta)\sin^3 (\alpha)  +2\kappa e^{-2R}\left(M_{\omega}^2 S^2 +K\bar{N}p^2 \sin^2(q \theta)\sin^2 (\alpha) \right)=0 \ . 
\end{gather}
Now, we can solve the equations for $\alpha$ and $R$ analytically, obtaining
\begin{gather}
    \alpha(X)=2\arctan\{ e^{qX+C_1} \} \ , \label{solalpha}\\ 
    R(X)=K \kappa \bar{N} \ln\{\cosh{(qX+C_1)}\}+C_2X+C_3 \ , \label{solR} 
\end{gather}
where $C_1$, $C_2$ and $C_3$ are integration constants.

Finally, using Eqs. \eqref{solalpha}  and \eqref{solR} it only remains to deal with the equations for $S$ (a Poisson equation with a mass term) and $H$, which can be easily integrated numerically. However, before doing this, it is convenient to study the appropriate boundary conditions that lead to topologically non-trivial solutions.

\subsection{Characterization of the solutions: Regularity, energy density  and baryon charge}

It is possible to check the regularity of the solutions presented above by computing the curvature invariants \cite{Sneddon}, \cite{Harvey}. As in Ref. \cite{gravtube1}, all the invariants are proportional to the Ricci scalar, which depends only on the metric function $R$ as
\begin{equation*}
\mathcal{R} =2 e^{2R}\partial_X^2 R \ .
\end{equation*}
As an example, the Kretschmann scalar, the square of the Ricci tensor, and the Weyl tensor invariants are the following:
\begin{equation*}
     \mathcal{K}\equiv  R^{\mu \nu\alpha\beta}{}_{\mu \nu\alpha\beta}= \mathcal{R}^2 \ , \quad 
    R_{\mu \nu} R^{\mu \nu} = \frac{1}{2} \mathcal{R}^2 \ , \qquad  C_{\mu \nu \lambda \rho} C^{\mu \nu \lambda \rho} = \frac{1}{3} \mathcal{R}^2 \ .
\end{equation*}
Even more, the function $R$ in Eq. \eqref{solR} is exactly the same as in the case of the gauged hadronic tubes presented in Ref. \cite{gravtube1}; therefore, the complete analysis of the regularity can be replicated. The key point is that the regular behavior of the solutions is determined by the soliton profile because the function $R$ is determined by $\alpha$, as can be seen from the Einstein equations, and does not depend in any way on the $\omega$-mesons. According to Ref. \cite{gravtube1}, regularity is ensured if the following condition is satisfied: 
\begin{equation}  \label{constraint}
    -(1-\bar{N}K\kappa) |q| < C_2 < - \bar{N} K \kappa |q| \ , 
\end{equation}
where the constants $C_1$ and $C_2$ have been introduced in Eqs. \eqref{solalpha} and \eqref{solR}. The above is the generalization of the constraint found in Ref. \cite{gravtube1} for the $SU(2)$ case, now for arbitrary flavors.

These conditions can also be read from the expression for the energy density ($\mathcal{E}= T_{tt}$), which is given by 

\begin{align}\notag
    \mathcal{E} \, = \,  & e^{2R}\left(\partial S\right)^{2}+M_{\omega}^{2}S^{2}-24\bar{N}e^{2R}\gamma pq\sin\left(q\theta\right)\sin^{2}(\alpha) \alpha' S
    \\&+\bar{N}K\left\{p^{2}\sin^{2}\left(q\theta\right)\sin^{2}\left(\alpha\right)+\frac{1}{2}e^{2R}(2-H)\left(q^{2}\sin^{2}\left(\alpha\right)+(\alpha')^{2}\right)\right\} \ .
\end{align}

Now, we can compute the topological charge for the matter field in Eqs. \eqref{U} and \eqref{dof} using the formula in Eq. \eqref{B}. The topological charge density is 
\begin{equation}
    \sqrt{-\sigma} \, n_0\rho^0 = 12 p q \bar{N} \sin(q\theta) \sin^2(\alpha) \, \alpha'  \ . \label{rho0}
\end{equation}
It is important to note that the integration can be performed by imposing boundary conditions for the soliton profile without the need to do the integration in the coordinate $X$ (since we can replace $\alpha' dX= d\alpha$). The periodicity of the $U$ matrix demands the following boundary condition for the $\alpha$ profile: 
\begin{equation}
    \alpha(\infty) - \alpha(-\infty) = 2 \pi n \ , 
\end{equation}
where $n$ is an integer. Using the above, we can see that for the topological charge to be a nonzero number, the constant $q$ must necessarily be $q= \frac{2m+1}{2}$, with $m$ an integer. Then, integrating over the spatial coordinates $X$ and $\theta$, we obtain the following expression for the topological charge per unit length of the tube,  
\begin{equation}
    \frac{B}{L_z} = \bar{N} n p  \ ,
\end{equation}
where $L_z = \frac{1}{\pi} \int dz$. Thus, we conclude that first, the baryon number is non-zero and proportional to the number of flavors if the constant $q$ is determined by the condition written above, and second, the baryon charge can be arbitrarily high due to the presence of the $n$ parameter allowed by the boundary conditions.

It is important to clarify the periodicity of $U$. While the matter ansatz \eqref{dof} implies that the matrix $U$ is not $2\pi$-periodic in the angular coordinate $\theta$, it is essential to note that both the spacetime metric and the energy density of the solution remain $2\pi$-periodic with respect to $\theta$ given the condition for $q$. Specifically, the angular dependence of the metric enters through the function $H$, which is itself periodic with a period of $2\pi$. Regarding the energy density, we observe that the angular dependence is determined by the $H$ and $S$ functions, along with a $\sin(q\theta)$ contribution. Given that the angular dependence of the meson potential also scales with $\sin(q\theta)$, the energy density remains periodic with a period of $2\pi$. Therefore, even though $U$ itself lacks $2\pi$-periodicity, the physical observables remain periodic, as the global periodicity can be restored via an isospin rotation

\begin{figure}[h!]
\begin{subfigure}[h!]{0.35\textwidth}
         \centering \includegraphics[width=1\textwidth]{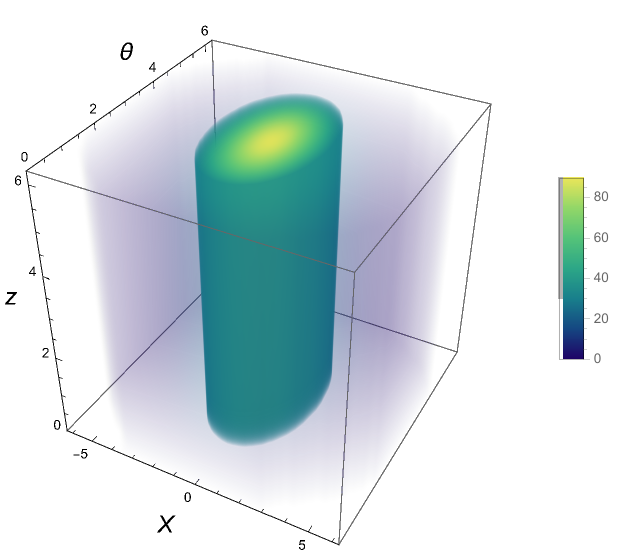}
         \caption{Case $SU(2)$}
\end{subfigure}
\begin{subfigure}[h!]{0.35\textwidth}
         \centering \includegraphics[width=1\textwidth]{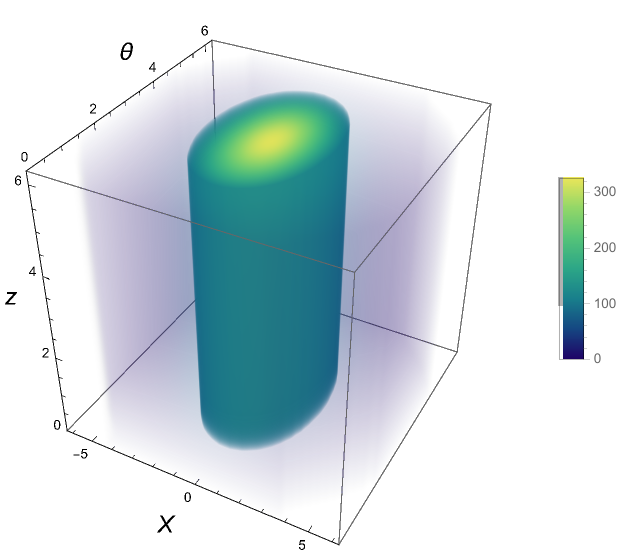}
         \caption{Case $SU(3)$}
\end{subfigure}\\
\begin{subfigure}{0.35\textwidth}
    \includegraphics[width=1\textwidth]{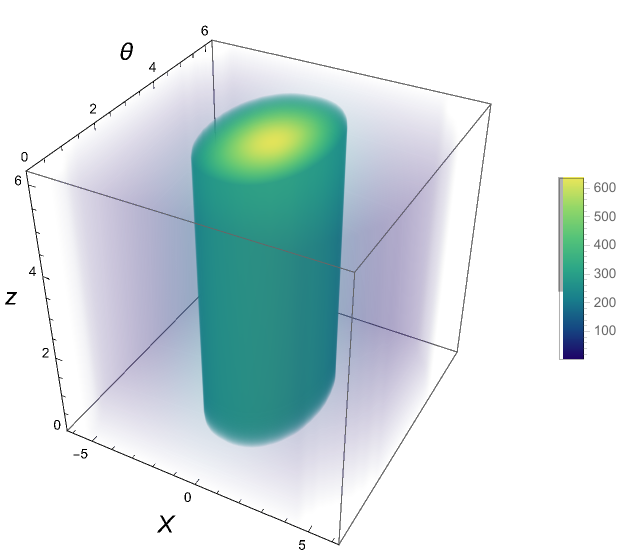}
    \caption{Case $SU(4)$}
\end{subfigure}
\caption{Energy density of the baryonic tube for different values of the flavor number. The presence of the $\omega$-mesons generates a flattening in one of the transverse directions. As the flavor number increases, the energy increases.}
\label{fig:comparison1}
\end{figure}

Fig. \ref{fig:comparison1} shows the energy density of the hadronic tubes for different values of the flavor number. For this purpose, we have assumed the following values for the free parameters: $ p=10, \, q=1/2, \, \gamma=0.01, \, C_1=0, \, C_2=-1/50, \, C_3=1$, and set the coupling constants as follows: $\kappa= 0.25, \, K=1, \,  M_\omega=0.1$.

We can see that, as in the flat space-time case in a finite-volume reported in Refs. \cite{ Barriga:2022izc} and \cite{Barriga:2023jam}, the presence of the $\omega$-mesons generates a flattening in one of the transverse directions. We also see that, as the flavor number increases, the energy density gradually becomes more dispersed.
The figures show that our solutions are regular, but it should be understood that the energy density is not located at the origin but is distributed around the axis. This becomes evident when one writes the solution in terms of the true radial coordinate $r$ in Eq. \eqref{int-r(X)}.

In line with \cite{gravtube1}, the solutions presented above correspond to cosmic tubes, as these are regular and free of singularities everywhere. Although these solutions do not present an angular deficit when the radial coordinate $r$ is close to the origin, asymptotically for $r\to\infty$ the solution presents an angular deficit, therefore they behave as a cosmic string at spatial infinity (for further details about the structure of cosmic strings see \cite{vilenkin1}-\cite{Moss:1987ka}).

\subsection{An analytical solution without $\omega$-mesons}

In the absence of $\omega$-mesons it is possible to solve the complete system of equations analytically. This scenario corresponds to the $SU(N)$-NLSM, for which the Einstein equations in Eq. \eqref{EqH} reduces to
\begin{gather}
            \partial^2_XH +\partial^2_\theta H  +2C_0 \sin^2(q \theta)\sin^2 (\alpha) e^{-2R}=0 \ , \label{eqHSinMesones}
\end{gather}
where $C_0=K \kappa\bar{N}p^2$. 
Using the identity; $2\sin ^2(q \theta)=(1-\cos (2 q \theta))$, Eq. \eqref{eqHSinMesones} can be written as
\begin{equation}
\Delta H=-F(X)(1- \cos (2 q \theta)) .\label{EqH_simplify}
\end{equation}
where we have defined $F(X)=C_0  e^{-2 R} \sin ^2(\alpha)$.
A natural Ansatz for the function $H(X,\theta)$ is
\begin{equation*}
    H(X,\theta)=H_1(X)+H_2(X)\cos{(2 q\theta) } \ . 
\end{equation*}
Replacing this Ansatz into Eq. \eqref{EqH_simplify}, we obtain the following ODE system,
\begin{gather*}
    H_1^{\prime\prime}=-F(X)\\
    H_2^{\prime\prime}-4 q^2H_2=F(X) \ . 
\end{gather*}
The first equation can be solved by quadratures,
\begin{equation}
    H_1(X)=-\int_{X_0}^X(X-s)F(s)ds \ , 
\end{equation}
while the second equation can be solved with the Green function's method; the corresponding solution is given by 
\begin{equation}
H_2(X)=\frac{1}{4 q} \int_{X_0}^X \sinh (2 q(X-s)) F(s) d s \ .
\end{equation}
Hence the general solution to Eq. \eqref{EqH_simplify} is given by
\begin{equation}
    H(X,\theta)=H_0(X,\theta)+H_1(X)+H_2(X)\cos(2 q\theta) \ , 
\end{equation}
where $H_0(X,\theta)$ is the solution to the homogeneous associated equation.

\section{The flat limit}

In this section, we discuss the flat limit of the above solutions by taking the limit $\kappa\to 0$. 

\subsection{Baryonic tubes at finite volume}

In the flat limit, one can check that the field equations for the metric functions, $R$ and $H$, become
\begin{gather}
    R^{\prime\prime}=0 \ , \label{EqRflat} \\ 
    \Delta H=0 \ , \label{EqHflat}
\end{gather}
while the matter field equations in Eqs. \eqref{Eqalpha} and \eqref{EqS} remain the same. It is also important to note that the first-order equation for the soliton profile, written in Eq. \eqref{Eqalpha0}, does not hold in the flat limit. This is because the equation is not exactly Eq. \eqref{Eqalpha0}, but rather
\begin{equation*}
K\kappa \bar{N} \, [ (\alpha')^{2}-q^{2}\sin\left(\alpha\right)^{2} ]=0 \ .
\end{equation*}
Now, Eq. \eqref{EqRflat} leads directly to
\begin{equation}
    R(X)=C_2 X+C_3  \ ,
\end{equation}
while the simplest solution for Eq. \eqref{EqHflat} (and which does not lead to a trivial solution of the system) is $H=0$.\footnote{Non-trivial solutions to Eq. \eqref{EqHflat} could lead to novel modulated baryonic configurations, which we hope to explore in a future publication.}
At this point, there are two possible paths to follow in the study of the flat limit, and these are determined by the cases $C_2=0$ and $C_2 \neq 0$.

When $C_2$ is nonzero, the geometry is reduced to that of a cylinder in which, to avoid the appearance of a conic singularity, the integration constant must necessarily be $C_2 = -1$. However, this is not the only issue to deal with. One can notice that in this case, the solution becomes singular because the energy density diverges at the origin. The only possible option in this scenario would be to take $q=0$ to avoid the divergence, but this condition is not compatible with the constraints that emerge for $q$ in the topological charge calculation presented in the previous section. Thus, the option to consider in the study of the flat limit is that with $C_2=0$. A detailed analysis of this procedure can be found in Ref. \cite{gravtube1}, Sec. VI.

Indeed, considering the case where $C_2=0$ and $C_3=0$ we reach the ``genuine flat-limit" of these solutions which, as expected, turns out to be regular. These configurations live in a flat space-time metric 
\begin{equation}
d s^2=-d \tilde{t}^2+d \tilde{z}^2+d X^2+  d \theta^2 \ ,
\end{equation}
where $    \tilde{t} = \frac{1}{\sqrt{2}} (2t -z)$, and $\tilde{z} = \frac{z}{\sqrt{2}}$. These configurations describe baryonic tubes confined in a finite volume. In these coordinates, the $\omega$-mesons potential and the degrees of freedom of the pionic field take the form 
\begin{gather}
    \alpha(x^\mu) = 2 \alpha(x) \ , \quad \Theta(x^\mu) = q y \ , \quad \Phi(x^\mu) = p(t-z) \ , \\ 
    \omega_\mu = (-S, 0, 0, S) \ , \qquad S=S(x,y) \ . 
\end{gather}
This kind of solution was obtained and studied in detail in Refs. \cite{ Barriga:2022izc} and \cite{Barriga:2023jam} (see also Refs. \cite{crystal1} and \cite{crystal2}).
Now, the present solution generalizes those reported in Refs. \cite{ Barriga:2022izc} and \cite{Barriga:2023jam} by including arbitrary flavors. In fact, the parameter $N$ appears explicitly in the equation for $S$:
\begin{equation} \label{EqSflat}
    \Delta S+12\bar{N}\gamma pq\sin\left(q\theta\right)\sin^{2}(\alpha)\alpha' -M_{\omega}^{2}S =0 \  .
\end{equation}
and, more importantly, in the expression for the energy density. The latter is given by 
\begin{equation} \label{EDflat}
    \varepsilon = (\partial S)^2  + \frac{1}{2}K \bar{N}\{ (q^2+2p^2\sin^2(q \theta))\sin^2(\alpha) +  (\alpha')^2 \} -24 \gamma \bar{N} pq \sin(q\theta) \sin^2(\alpha) S \alpha'  +M_\omega^2 S^2 = 0  \  . \end{equation}
Recalling that the equation for $\alpha$ in Eq. \eqref{Eqalpha} can be solved exactly in terms of elliptic functions, we see that the complete system of equations in the flat limit is reduced to a single equation for $S$ that can be approached using Green functions. Here, our interest is in calculating the binding energy, so we will simply solve Eq. \eqref{EqSflat} using numerical integration.

As we have pointed out previously, one of the benefits of introducing the $\omega$-mesons into the NLSM is that it allows us to reduce the predicted binding energy for nucleons, thus achieving estimates closer to the experimental data. Here, we will calculate the binding energy of the nucleons as a function of the flavor number. The binding energy can be estimated using the formula
\begin{equation} \label{binding}
    \Delta (B) = \frac{E_{(B+1)}-(E_{(B)}+E_{(1)})}{(B+1) E_{(1)}} \ ,
\end{equation}
where $B$ is the baryon number of the baryonic tubes and $E(i)$ is the total energy of the system that contains $(i)$ baryons obtained from the integration of Eq. \eqref{EDflat}.
\begin{figure}[h!]
\begin{subfigure}[h!]{0.35\textwidth}
         \centering \includegraphics[width=1.1\textwidth]{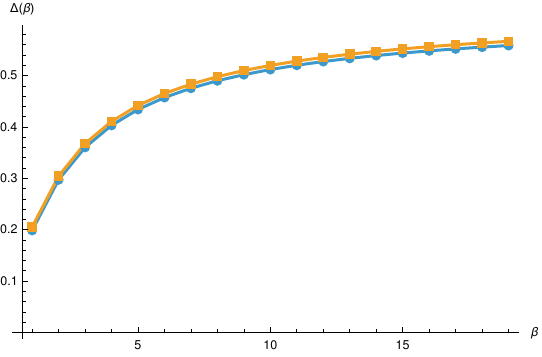}
         \caption{$SU(2)$}
\end{subfigure} \qquad \qquad
\begin{subfigure}[h!]{0.35\textwidth}
         \centering \includegraphics[width=1.1\textwidth]{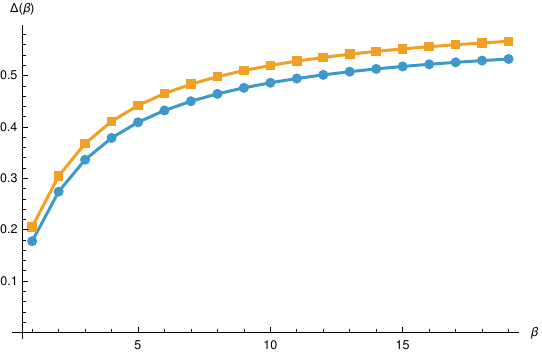}
         \caption{$SU(3)$}
\end{subfigure}\\
\begin{subfigure}{0.5\textwidth}
    \includegraphics[width=1.1\textwidth]{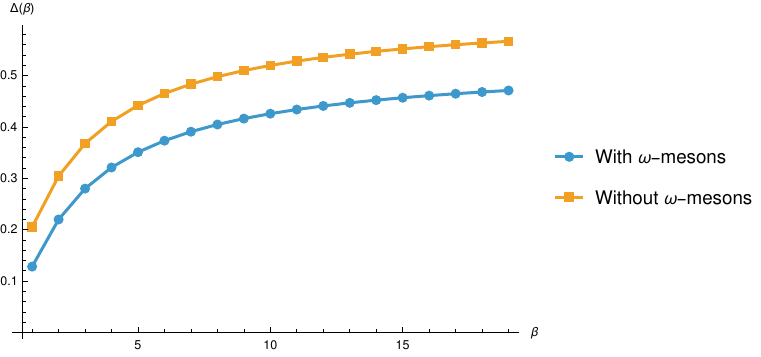}
    \caption{$SU(4)$}
\end{subfigure}
\caption{Binding energy as a function of the baryon number of a system of baryons at a finite volume. The inclusion of the $\omega$-mesons reduces the binding energy for any value of the flavor number.}
\label{fig:comparison2}
\end{figure}
\begin{figure}[h!]
\centering \includegraphics[width=.6\textwidth]{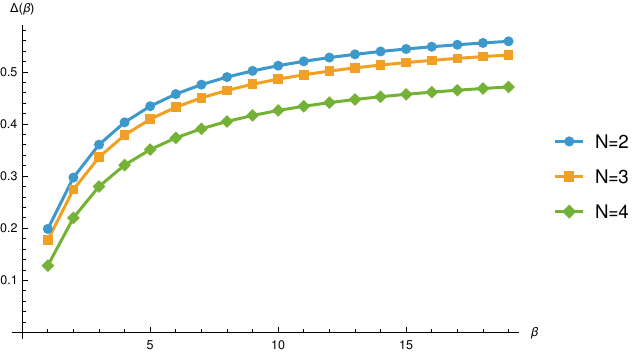}
\caption{Binding energy as a function of the flavor number for different values of $N$. As the number of flavors increases, the binding energy decreases.}
\label{fig:comparison3}
\end{figure}
Taking into account the solutions of Eqs. \eqref{Eqalpha} and \eqref{EqS}, and introducing this information into the expression for the energy density in Eq. \eqref{EDflat}, we obtain the plots of the binding energy shown in Fig. 2 and Fig. 3, where we have defined $\beta=B/\bar{N}$. In Fig. 2, we can see that the inclusion of the $\omega$-mesons in the NLSM allows the predicted binding energy for hadronic tubes to be reduced for any value of $N$. This behavior is consistent with that predicted in pioneering work in Ref. \cite{Adkins1}, and with the results derived later in Refs. \cite{Barriga:2023jam} and \cite{Barriga:2022izc}. Fig. 2 shows a novel comparison of the binding energy for different values of the number of flavors. Here we can see that, as the number of flavors increases, the binding energy becomes progressively smaller. To the authors' knowledge, this fact had not been explicitly demonstrated before, and it underscores the importance of including more than two flavors in effective models to make them compatible with experimental data. For the plots, we have set $K=1$, $\gamma=0.04$, $p=q=1$ and $M_\omega=0.01$.

\subsection{On the emergent chemical potential}

At this point, it is important to highlight a particular feature related to the time dependence of the matter fields. According to the Ansatz introduced in Eq. \eqref{dof}, we observe that although the fields are time-dependent, both the energy density and the topological charge density remain time-independent (even for the self-gravitating configuration); thus, the configurations presented here are strictly static. Furthermore, the time dependence in our Ansatz can be interpreted as an emergent isospin chemical potential (where the time dependence represents a uniform rotation in the isospin space), but only in the sector without $\omega$-mesons, as explained below.

The standard way to introduce an isospin chemical potential, $\mu_I$, in the context of effective field theories, is through the covariant derivative 
\begin{equation*}
    \partial_{\mu} U \rightarrow \partial_{\mu} U + \mu_{I}\left[T_3, U \right]g_{\mu t} \ . 
\end{equation*}
Then, considering the following static Ansatz for the matter field:
\begin{gather} \label{staticAnsatz}
    \alpha(x^\mu) = \alpha(x) \ , \quad \Theta(x^\mu) = q y \ , \quad \Phi(x^\mu) = pz \ , \\ 
    \omega_\mu = (-S, 0, 0, S) \ , \qquad S=S(x,y) \ , 
\end{gather}
in the presence of the isospin chemical potential defined above, it can be directly verified that the equations of the NLSM (without $\omega$-mesons) are still reduced to the equations for the profile shown in Eq. \eqref{Eqalpha}. The price to pay for obtaining exactly the same equation (and therefore the exact same solution) is the fixing of the chemical potential to a particular value, namely, $\mu_I= p$. This establishes a direct relation between the isospin and the baryon number of the configurations. Indeed, in flat space-time, by imposing the boundary conditions $\alpha(2 \pi) - \alpha(0) = n \pi$, the topological charge becomes $B=\bar{N} n p$. From the above, we can conclude that the time dependence in our Ansatz can be interpreted as the appearance of an isospin chemical potential in the NLSM, which is finite and directly proportional to the baryon number. 

However, with the inclusion of the $\omega$-mesons, this construction is no longer possible; specifically, for the Ansatz in Eq. \eqref{staticAnsatz}, there is no value of the isospin chemical potential that leads to Eqs. \eqref{Eqalpha} and \eqref{EqSflat}. This issue can be directly verified, as the use of the static Ansatz leads to mathematical inconsistencies at the level of the field equations that prevent their solution. There is a clear technical reason behind this: the coupling of the NLSM with the $\omega$-mesons occurs at the level of the action through the interaction of the vector field with the topological current. This contribution breaks the symmetry of the NLSM in the isospin space. By introducing the isospin chemical potential as a component of an external gauge field, it only affects the terms containing covariant isospin derivatives (the NLSM sector) but not the interaction term $\omega_\mu \rho^\mu$ in Eq. \eqref{I}, since the $\omega$-meson is an isosinglet.

In conclusion, the equivalence between time dependence and an isospin chemical potential only holds if the entire dynamics of the system are sensitive to the phase transformation in the same manner. The $\omega$-meson, being a massive field that acts as a stabilizer (providing short-range repulsion), breaks the ``substitution symmetry" that the NLSM possesses on its own.

\section{Conclusions and perspectives}

In this paper, we have shown that the Einstein-NLSM in $D=4$ spacetime dimensions coupled to $\omega$-mesons admits topological soliton solutions describing self-gravitating baryon tubes. Furthermore, by employing the maximal embedding Ansatz of $SU(2)$ into $SU(N)$ in the exponential representation, these solutions have been proven to be robust for an arbitrary number of flavors. These baryon tubes do not exhibit singularities, and their topological charge is proportional to the number of flavors in the theory. 

The flat limit of these solutions is particularly interesting, as the complete system of equations reduces to a single equation for the $\omega$-meson potential, which can be solved in terms of Green's functions. These new solutions generalize those previously found in Refs. \cite{Barriga:2022izc} and \cite{Barriga:2023jam} for the two-flavor case. As it is well-known, the coupling of $\omega$-mesons to the NLSM allows for a reduction in the repulsion energy between baryons, resulting in theoretical estimates that are closer to experimental data. In the present work, we have demonstrated this phenomenon by comparing the energy of the hadronic tubes with and without the presence of $\omega$-mesons. We have also observed that as the number of flavors in the theory increases, the binding energy decreases, suggesting that models with a higher number of flavors provide more accurate and physically realistic predictions. A key distinction between these two methods of reducing the binding energy is that, when coupling with $\omega$-mesons, the topological charge remains unaffected, which is a highly desirable feature.

This study opens promising directions for further investigation. For instance, the inclusion of large $N_c$ corrections and the coupling of the theory to Maxwell electrodynamics are natural extensions. It will also be important to analyze how these solutions are modified in the presence of a non-zero cosmological constant. Additionally, since it is known that the families of WLP solutions are related to the Kasner parameter \cite{Peraza:2025}, it would be of great interest to explore how the number of flavors influences this parameter.

\acknowledgments

The authors thank Fabrizio Canfora, Cristóbal Corral, Marcela Lagos, and Javier Peraza for their useful comments. G.B. is funded by the National Agency for Research and Development ANID grant 21222098. C. H appreciates the support of FONDECYT postdoctoral Grant 3240632. A. V.
has been funded by FONDECYT Iniciaci\'on No. 11261883. L. S. is
grateful to Luca Ciambelli for his hospitality at Perimeter Institute during this work.
 L. S. is supported by Beca Doctorado Nacional ANID Grant No. 21221813 and the ANID Complementary Benefits for Research Internships. This research was supported in part by the Perimeter Institute
for Theoretical Physics. Research at Perimeter Institute is supported by the Government of Canada through the
Department of Innovation, Science, and Economic Development, and by the Province of Ontario through the
Ministry of Colleges and Universities.

\appendix

\section{About the radial coordinate}  \label{A3}

As we have pointed out previously, the coordinate $X$ in Eq.  \eqref{metric} is not strictly a radial coordinate, since it can, in principle, take both positive and negative values. An appropriate radial coordinate can now be defined as follows:
\begin{equation}
    r(X)=\int_{-\infty}^{r}e^{-R(y)}dy= \int_{-\infty}^{r}\frac{e^{-C2y -C_{3}}}{\left(\cosh(q y +C_{1}) \right)^{K \kappa}} dy \ , \label{int-r(X)}
\end{equation}
where $R(y)$ is given by the solution presented in Eq. \eqref{solR}.
One of the advantages of the function defined above is that $r(X)$ increases as $X$ increases; this is
\begin{eqnarray*}
    \frac{dr}{dX}=\frac{e^{-C_{2}X -C_{3}}}{\left(\cosh(q X +C_{1}) \right)^{K\kappa }} >0  \ .
\end{eqnarray*}
After we perform the integration of the expression on \eqref{int-r(X)}, we see that to have $r$ well-defined, that is, to be positive, it is necessary to impose the following condition on the integration constant  
\begin{eqnarray*}
 C_{2}<\kappa K N |q|  \ .
\end{eqnarray*}
When the above condition is considered, we see that 
\begin{eqnarray*}
    \lim_{X \rightarrow -\infty} r(X)=0  \ ,
\end{eqnarray*}
and therefore
\begin{eqnarray*}
    -\infty < X < \infty \Rightarrow 0 < r < \infty \ .
\end{eqnarray*}
Under that transformation, the metric takes the form 
\begin{equation}
ds^{2} =  \left(-2+ H\right)dt^{2} +2 \left(1-H\right) dtdz+H dz^2 + dr^2+e^{-2\tilde{R}}d\theta^{2}\ ,
\label{metricnew}
\end{equation}
where the function becomes $\tilde{R}(r)=R\left(X(r)\right)$\,. 

It is important to determine whether our solution exhibits an angular deficit. Since the transverse section of our metric is identical to one used in Ref. \cite{gravtube1}, it is possible to see that for a suitable value of the constant $C_{3}$ in the function $\tilde{R}(r)$, the solution is free of angular deficit when $r$ is close to zero, but develops an angular deficit at large $r$. Thus, our solution behaves asymptotically as a cosmic string. 


\end{document}